\documentstyle[epsf,psfig,proceedings,numreferences]{maxent95} 

\begin{opening}
\title{Maximum-likelihood method in quantum estimation}
\author{M. G. A. Paris, G. M. D'Ariano and M. F. Sacchi}
\institute{Quantum Optics Group, Unit\`a INFM and Dipartimento 
``A. Volta'' \\ Universit\`a di Pavia, via Bassi 6, I-27100 Pavia, 
ITALY\footnote{Email: paris@unipv.it}}
\end{opening}
\runningtitle{Maxlik in quantum estimation}
\begin{document}
\begin{abstract}  
The maximum-likelihood method for quantum estimation is reviewed 
and applied to the reconstruction of density matrix of spin and 
radiation as well as to the determination of several parameters 
of interest in quantum optics. 
\end{abstract}
\section{Introduction}
Quantum estimation of states, observables and parameters is, from
very basic principles, matter of statistical inference after sampling
a population. Why is it so ? And what does this statement exactly mean ?
The quantum description of a physical system is intrinsically a statistical 
one since given an unknown quantum state 
$\hat\varrho \,$: i) one cannot determine the quantum state from a single
measurement (even joint measurement) performed on the system \cite{yuend}; 
ii) it is not possible to measure different observables after copying
 the state preparation at disposal, since the linearity of quantum 
mechanics leads to the impossibility of cloning an arbitrary state \cite{wot}. 
In a way, this means that any quantum estimation procedure necessary 
requires many identical preparations of the system under examination, on which
one performs repeated measurements of an observable or of a set of 
observables. \par
The full inference of the quantum state from feasible measurements is 
a hot topic of the last decade. In recent years, experiments have been 
performed to characterize the quantum state of different physical systems,  
such as single-mode radiation field \cite{QHT}, 
a diatomic molecule \cite{Molecule}, 
a trapped ion \cite{Ion}, and an atomic beam \cite{Beam}. 
These results stimulated some efforts for efficient data-processing 
algorithms, in order to extract  the maximum available information on the quantum state,
since one always deals with finite ensembles 
\cite{FiniteEnsembles}, and every detection scheme is affected by noise and 
imperfections. The need for an efficient data 
processing is even more stringent in cases where one is not interested in the 
complete characterization of the state, but only in some specific feature, as when 
one addresses the problem of characterizing a device, rather than a quantum 
state, as, for example, the estimation of  coupling constants, gain
coefficients, or nonlinear susceptibilities. \par
The most comprehensive quantum estimation procedure is quantum
tomography \cite{general}. In quantum tomography the expectation value of
an operator is obtained by averaging a special function (usually termed
{\em sampling kernel} or {\em pattern function}) over the experimental 
data of a sufficiently complete set of observables which is called a 
``quorum''. For example, in homodyne tomography of radiation the quorum 
observables are the quadratures of the e.m. field (for varying phase with 
respect to the local oscillator). The expectation value of a generic operator 
is obtained by averaging the corresponding pattern function over data. 
The method is very general and efficient, however, in the averaging procedure, 
we have fluctuations which result in large statistical errors. \par
In this paper we review the the maximum-likelihood (ML) principle 
approach to the 
quantum estimation problem. The ML idea is to find the quantum state, or the 
value of the parameters, that are most likely to generate the observed data. 
This idea can be quantified and implemented using the concept of the 
likelihood functional. Concerning the estimation of quantum state, 
in contrast to quantum tomography, the ML method estimates the state as 
a whole. As a result, {\em a priori} knowledge about properties of the density 
matrix can be incorporated from the very beginning, thus assuring positivity 
and normalization of matrix, with the result of a substantial
reduction of statistical errors \cite{maxlik}. Regarding the estimation of specific 
parameters, we notice that in all the cases here analyzed the resulting 
estimators are efficient, unbiased and 
consistent, thus providing a statistically reliable determination 
\cite{parlik}. Moreover, by 
using the ML method only small samples of data are required for a precise 
determination.  
\par\section{Maximum likelihood principle} 
Here we briefly review the theory of the maximum-likelihood (ML)
estimation of a single parameter. The generalization to several
parameters, as for example the elements of the density matrix is, in 
principle, straightforward. The only point that should be carefully 
analyzed is the parameterization of the multidimensional quantity to 
be estimated. In the next section the specific case of the density matrix
will be discussed. \par 
Let $p(x | \lambda)$ the probability density of a random variable $x$, 
conditioned to the value of the
parameter $\lambda$.  The form of $p$ is known, but the true value of
the parameter $\lambda$ is unknown, and will be estimated from the
result of a measurement of $x$.  Let $x_1, x_2, ..., x_N$ be a random
sample of size $N$. The joint probability density of the independent
random variable $x_1, x_2, ..., x_N$ (the global probability of the
sample) is given by
\begin{eqnarray}
{\cal L}(x_1, x_2, ..., x_N| \lambda)= \Pi_{k=1}^N \: p(x_k |\lambda)
\label{likdef}\;,
\end{eqnarray}
and is called the likelihood function of the given data sample
(hereafter we will suppress the dependence of ${\cal L}$ on the data). The
maximum-likelihood estimator (MLE) of the parameter 
$\lambda$ is defined as the quantity $\lambda_{\sc ml} \equiv \lambda_{\sc ml} 
(\{x_k\})$ that maximizes ${\cal L} 
(\lambda)$ for variations of $\lambda$, namely $\lambda_{\sc ml}$ is 
given by the solution of the equations
\begin{eqnarray}
\frac{\partial {\cal L} (\lambda) } {\partial \lambda} = 0 \; ; \quad 
\frac{\partial^2 {\cal L} (\lambda)} {\partial \lambda^2} < 0
\label{maxlikdef}\;.
\end{eqnarray}
Since the likelihood is positive the first equation is
equivalent to $\partial L/ \partial \lambda = 0 $ 
where 
\begin{eqnarray}
L(\lambda) = \log {\cal L} (\lambda) = \sum_{k=1}^N \log p(x_k | \lambda)
\label{loglikfun}\;
\end{eqnarray}
is the so-called log-likelihood function. 
\par In order to obtain a measure for the confidence interval in  
the determination of $\lambda_{\sc ml}$ we consider the variance
\begin{eqnarray}
\sigma^2_\lambda = \int \left[\prod_k dx_k \: p(x_k|\lambda)\right] 
\left[\lambda_{\sc ml} 
(\{x_k\})- \lambda \right]^2 \: \label{varMLEdef}\;.
\end{eqnarray}
Upon defining the Fisher information 
\begin{eqnarray}
F= \int dx \left[ \frac{\partial p(x |\lambda)}{\partial \lambda}\right]^2
\frac1{p(x | \lambda)}
\label{FisherDef}\;,
\end{eqnarray}
it is easy to prove \cite{tucker} that 
\begin{eqnarray}
\sigma^2_\lambda  \geq (N\:F)^{-1}
\;\label{manca}
\end{eqnarray}
where $N$ is the number of measurements. The inequality in
Eq. (\ref{manca}) is known as 
the Cram\'er-Rao bound \cite{cramer}
on the precision of ML estimation. Notice that this bound holds 
for any functional form of the probability distribution 
$p(x|\lambda)$, provided that the Fisher information exists $\forall \lambda$
and $\partial_\lambda p(x|\lambda)$ exists $\forall x$. When an experiment 
has "good statistics" (i.e. a data sample large enough) the Cram\'er-Rao 
bound is saturated.
\par\section{Quantum state estimation}
In this section we review the the method for the 
maximum likelihood estimation of the quantum state, focusing attention
to the cases of homodyne and spin tomographies \cite{maxlik}. 
The physical situation we have in mind is an experiment consisting 
of $N$ measurements performed on identically prepared copies of a given
system. Quantum mechanically, each measurement is described by a
positive operator-valued measure (POVM). The outcome of the $i$th
measurement corresponds to the realization of a specific element of
the POVM used in the corresponding run. We denote this element
by $\hat\Pi_i$.  The likelihood is here a functional of the density matrix
${\cal L}(\hat{\varrho})$ and is given by the
product \begin{equation} {\cal L}(\hat{\varrho}) = \prod_{i=1}^{N}
\hbox{Tr}(\hat{\varrho} \hat\Pi_i)\:,  \end{equation} 
which represents the probability of the observed data.
The unknown element of the above expression, which we want to infer from 
data, is the density matrix describing the measured ensemble. \par
We restrict ourselves to finite dimensional Hilbert spaces. In this case, it can be 
proved that ${\cal L}(\hat{\varrho})$ is a concave function defined on a convex and 
closed set of density matrices.  Therefore, its maximum is achieved either on a 
single isolated point, or on a convex subset of density matrices. 
In the first case we have a proper reconstruction scheme, namely the set
of observables chosen for the measurement provides the complete
characterization of the state under examinations. On the other hand, if the
maximum is not unique, the set of
chosen observables is insufficient, namely it  
does not constitute a quorum.  
\par In order to optimize the procedure for the maximization of the 
likelihood function we introduce a specific parameterization of the set 
of density matrices. A given density matrix can be written in the form
\begin{equation}
\label{Eq:rhoTT}
\hat{\varrho} = \hat{T}^{\dagger} \hat{T}\;,
\end{equation}
which automatically guarantees that $\hat{\varrho}$ is positive 
and Hermitian for any complex lower triangular matrix $\hat T$, 
with real elements on the diagonal. 
For an $M$-dimensional Hilbert space, the number of
real parameters in the matrix $\hat{T}$ is $M+2M(M-1)/2=M^2$, which
equals the number of independent real parameters for a Hermitian
matrix. This confirms that our parameterization is minimal, up to the
unit trace condition. \par
In numerical calculations, it is convenient to replace the likelihood
functional by its natural logarithm, which does not change the
location of the maximum. Thus the function subjected to numerical
maximization is given by
\begin{equation}
\label{eq:lt}
L(\hat{T}) = \sum_{i=1}^{N} \ln \hbox{Tr}(\hat{T}^\dagger
\hat{T} \hat\Pi_i) - \lambda \hbox{Tr}(\hat{T}^\dagger
\hat{T})\;,
\label{loglik}
\end{equation}
where $\lambda$ is a Lagrange multiplier accounting for normalization
of $\hat \varrho$ that equals the total number of measurements
$N$ \cite{not}. 
Using this formulation, the maximization problem can be solved by
standard numerical procedures for searching the maximum over the 
$M^2$ real parameters of the matrix $\hat{T}$ \cite{maxlik}. The examples
presented below use the downhill simplex method \cite{Ameba}. \par
Our first example is the application of the ML estimation in quantum
homodyne tomography of a single-mode radiation field \cite{DLP}, which
is so far the most successful method in measuring nonclassical states
of light \cite{QHT,noncl}. The experimental apparatus used in this
technique is the homodyne detector.  The realistic, imperfect homodyne
measurement is described by the positive operator-valued measure
\begin{equation} 
\label{Eq:Hxphi}
\hat{\cal H}(x;
\varphi) = \frac{1}{\sqrt{\pi(1-\eta)}} \exp \left( - \frac{(x-\sqrt{\eta}
\hat{x}_{\varphi})^2}{1-\eta}\right)\:, 
\end{equation}
where $\eta$ is the detector efficiency, and  $\hat{x}_\varphi
=(a^\dag \,e^{i\varphi}+a\,e^{-i\varphi})/2$ is
the quadrature operator ($[a,a^\dag ]=1$), 
depending on the externally adjustable
local oscillator (LO) phase $\varphi$. \par
After repeating the measurement $N$ times, we obtain a set of pairs
$(x_i; \varphi_i)$ consisting of the outcome $x_i$ and the LO phase
$\varphi_i$ for the $i$th run, where $i=1,\ldots, N$. The
log-likelihood functional is given by Eq. (\ref{loglik}) with
$\hat\Pi_i\equiv\hat{\cal H}(x_i; \varphi_i)$.  Of course, for a
light mode it is necessary to truncate the Hilbert space to a finite
dimensional basis. We shall assume that the highest Fock state has
$M-1$ photons, i.e.\ that the dimension of the truncated Hilbert space
is $M$.  For the expectation $\mbox{Tr}[\hat{T}^{
\dagger}\hat{T}\hat{\cal H}(x;\varphi)]$ it is necessary to use 
an expression which is explicitly positive, in order to  protect the algorithm
against occurrence of small negative numerical arguments of the
logarithm function. A simple derivation yields
\begin{eqnarray}
\mbox{Tr}[\hat{T}^{\dagger}\hat{T}\hat{\cal
H}(x; \varphi)] 
 =  \sum_{k=0}^{M-1}\sum_{j=0}^{k}
\left|\sum_{n=0}^{k-j} 
\langle k | \hat{T} | n+j \rangle B_{n+j,n} \langle n |
x\rangle e^{in\varphi}\right|^2 \;,
\end{eqnarray}
where $B_{n+j,n} = \left[{{n+j} \choose n} \eta^{n}
(1-\eta)^{j}\right]^{1/2} $ and $ \langle n | x \rangle = H_n(x)
\exp(-x^2/2) / \sqrt{2^n n! \pi^{1/2}} $ are eigenstates of the
harmonic oscillator in the position representation---$H_n (x)$ being
the $n$th Hermite polynomial.\par
\begin{figure}[h] \begin{center}\vspace{-26pt}
\epsfxsize=.45\hsize\leavevmode\epsffile{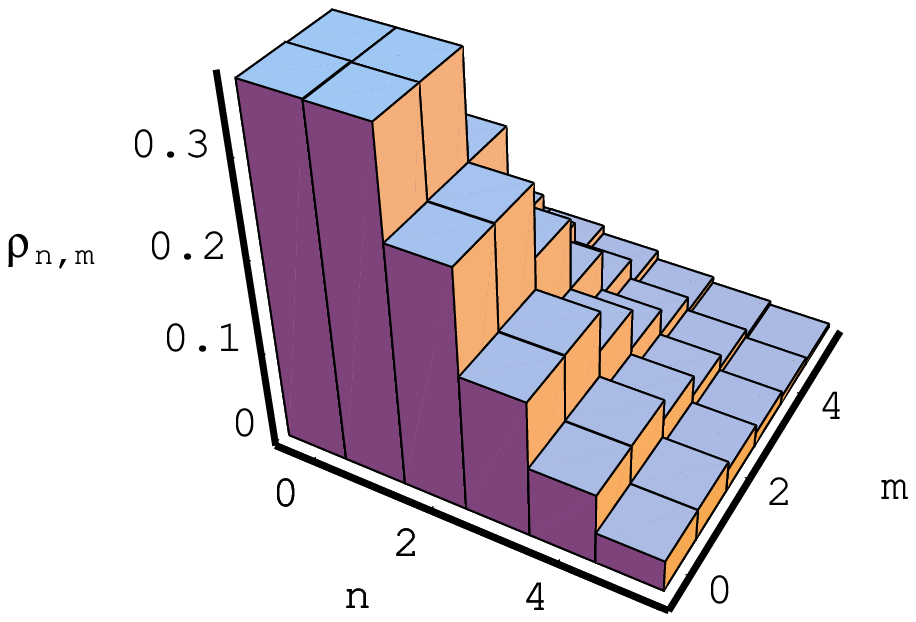}
\epsfxsize=.45\hsize\leavevmode\epsffile{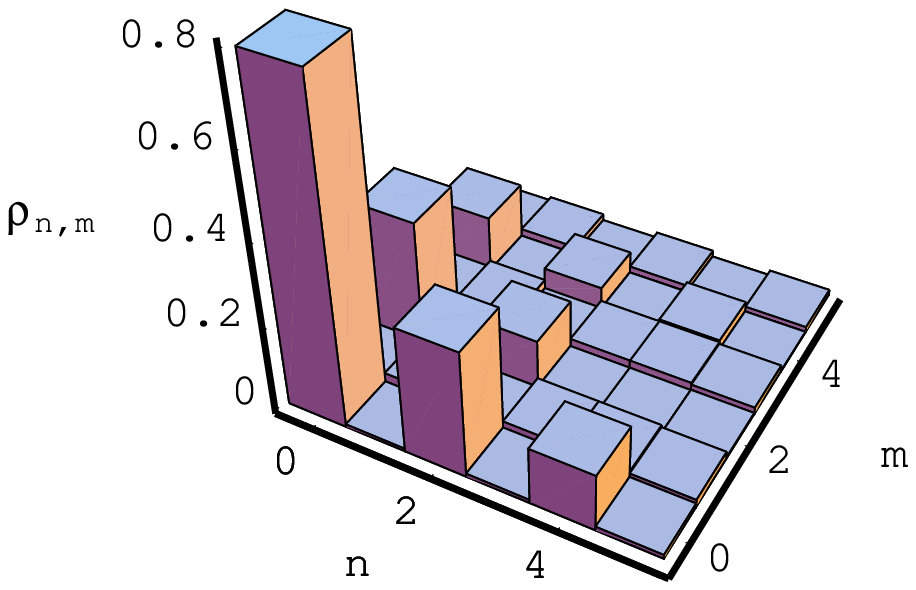}
\caption{Reconstruction of the density matrix of a single-mode
radiation field by the ML method. The plot shows the matrix elements 
of a coherent state (left) with $\langle \hat a^{\dag }\hat a \rangle =1$
photon, and for a squeezed vacuum (right) with $\langle \hat a^{\dag
}\hat a \rangle =0.5$ photon. A sample of 50000 simulated homodyne 
data for quantum efficiency $\eta=80\%$ has been used.}\label{Fig:QHT}
\end{center} \end{figure}

We have applied the ML technique to reconstruct the
density matrix in the Fock basis from Monte Carlo simulated homodyne
statistics \cite{maxlik}.  Fig.~\ref{Fig:QHT} depicts the matrix elements of the
density operator as obtained for a coherent state and a squeezed
vacuum. Remarkably, only 50000 homodyne data have been
used for quantum efficiency at photodetectors $\eta=80\%$.
We notice that the ML method is affected by much smaller statistical 
errors than conventional tomography. As a comparison one could see 
that the same precision of the reconstructions in Fig.~\ref{Fig:QHT} 
could be achieved using $10^7$--$10^8$ data samples in conventional tomography
of Ref.~\cite{DLP}.  On the other hand, in order to find numerically the ML
estimate we need to set {\em a priori} the cut-off parameter for the
photon number, and its value is limited by increasing computation
time. 
\par\vspace{-13pt}
We mention that ML estimation can also be applied to the reconstruction 
of the quantum state a two-mode field \cite{maxlik}, along with the multi-mode
tomographic technique with a single LO \cite{single}.
\par
\begin{figure}[h]\begin{center}
\epsfxsize=.5\hsize\leavevmode\epsffile{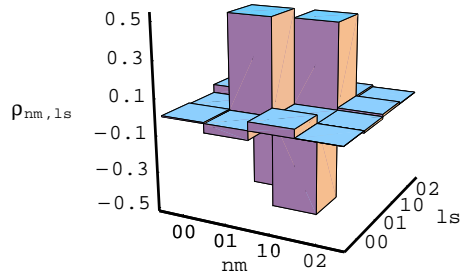} \end{center}
\caption{Reconstruction of the 
density matrix of a pair of spin-1/2
particles in the singlet state by ML method.  The matrix elements has been 
obtained by a sample of 500 simulated data. \label{Fig:singlet}}
\end{figure}

 Finally, we apply the ML procedure 
for reconstructing the density matrix of spin systems. For example,
let us consider 
$N$ repeated preparations of a pair of spin-1/2
particles. The particles are shared by two parties. In each run, the
parties select randomly and independently from each other a direction
along which they perform spin measurement.
The obtained result is described by the joint
projection operator over spin coherent states
\begin{equation}
\hat{\cal F}_i = |\Omega^A_{i}, \Omega^B_{i} \rangle \langle
\Omega^A_{i}, \Omega^B_{i} |\;,
\end{equation}
where $\Omega^A_{i}$ and $\Omega^B_{i}$ are the vectors on the Bloch sphere
corresponding to the outcomes of the $i$th run, and the indices
$A$ and $B$ refer to the two particles. 
As in the previous examples, it is convenient to 
use an expression for the quantum expectation value $\mbox{Tr}(\hat{T}^{
\dagger}\hat{T}\hat{\cal F}_i$) which is explicitly positive. 
The suitable form is 
$$\mbox{Tr}(\hat{T}^{\dagger} \hat{T}\hat{\cal F}_i)= \sum_\mu 
|\langle \mu | \hat{T} | \Omega^A_{i}, \Omega^B_{i} \rangle|^2\;,$$
where $|\mu\rangle$ is an orthonormal basis in the Hilbert space
of the two particles. The result of a simulated experiment with only
500 data for the
reconstruction of the density matrix of the singlet state is shown in
Fig. \ref{Fig:singlet}.
\par\section{Parameters estimation in quantum optics}
Here we focus our attention on the determination of 
specific parameters which are relevant in quantum optics, and analyze 
their ML estimation procedure in some details. In the next two subsections
we consider the estimation of the parameters of a Gaussian state and 
the estimation of the quantum efficiency of both linear and avalanche 
photodetectors. The reader may found more details in Ref. \cite{parlik}.
\par\subsection{Gaussian state estimation}
In this section we apply the ML method to estimate the quantum state 
of a single-mode radiation field that is characterized by a Gaussian 
Wigner function. Such kind of states comprises the wide class of 
coherent, squeezed and thermal states, namely most of the states effectively
produced in an optical laboratory. We consider the Wigner function of the 
form 
\begin{eqnarray}
W(x,y)=\frac{2\Delta ^2}{\pi}\exp\left\{-2\Delta^2\left[\frac 1\kappa 
(x-a)^2 +\kappa (y-b)^2\right]\right\}\;,\label{wxy}
\end{eqnarray}
and we apply the ML technique starting from homodyne detection 
to estimate the four real parameters 
$\Delta , \kappa, a$ and $b$. 
The four parameters are connected to the number of thermal, squeezing and 
coherent-signal photons in the quantum state as follows
\begin{eqnarray}
n_{th}=\frac 12\left(\frac {1}{\Delta ^2}-1\right)\qquad 
n_{sq}=\frac{1+\kappa^2}{4\kappa}-\frac12 \qquad n_{coh}=a^2+b^2\;.
\end{eqnarray}
In fact, 
the quantum state corresponding to the Wigner function in Eq. (\ref{wxy}) 
writes
\begin{eqnarray}
\hat \varrho =D(\mu)\,S(r)\,\frac
{1}{n_{th}+1}\left(\frac{n_{th}}{n_{th}+1}\right)^{a^\dag a}
\,S^\dag (r)\,D^\dag (\mu)\;,
\end{eqnarray}
with $r=\frac 12 \log\kappa $ and $\mu =a+i\,b$, and 
where $S(r)=\exp[r(a^2-a^{\dag 2})/2]$ 
and $D(\mu)=\exp(\mu a^\dag -\mu ^*a)$ denote the squeezing 
and displacement operators, respectively. 
\par We consider repeated preparations of a Gaussian state, on which we
perform homodyne measurements at different phases $\phi$ with respect to 
the local oscillator. The homodyne distribution is given, for unit quantum 
efficiency of photodetectors, by the Gaussian \cite{yuen} 
\begin{eqnarray}
p(x,\phi)=\sqrt{\frac{2 \Delta ^2\kappa}{\pi(\kappa^2\cos^2\phi
+\sin^2\phi)}}\exp\left\{-2 \Delta ^2\kappa\frac{\left[x-a\cos\phi
-b\sin\phi)\right]^2}{\kappa^2\cos^2\phi+\sin^2\phi}\right\}
\;.\label{pxfi}
\end{eqnarray}
For non-unit quantum efficiency the ideal distribution of Eq. (\ref{pxfi})
is replaced by a convolution with a Gaussian of variance $(1-\eta)/(4\eta)$.
From Eqs. (\ref{loglikfun}) and (\ref{pxfi}) one easily evaluates 
the log-likelihood
function for a set of $N$ homodyne outcomes $x_i$ at random phases 
$\phi _i$ as follows
\begin{eqnarray}
L=\sum_{i=1}^N \frac12\log \frac{2\Delta^2\kappa}{\pi 
(\kappa^2\cos^2 \phi_i+\sin^2\phi_i)}-
2 \Delta ^2\kappa\frac{\left[x_i-a\cos\phi-b\sin\phi )\right]^2}
{\kappa^2\cos^2\phi_i+\sin^2\phi_i}
\;.\label{lgau}
\end{eqnarray}
The ML estimators $\Delta _{\sc ml}, \kappa_{\sc ml}, a_{\sc ml}$ 
and $b_{\sc ml}$ are found upon maximizing Eq. (\ref{lgau}) 
versus $\Delta, \kappa, a$ and $b$.
\par
\begin{figure}[h] 
\begin{center}
\centerline{\psfig{file=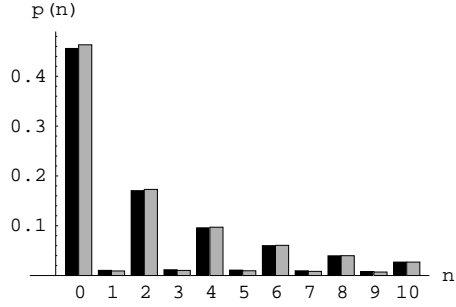,width=6cm}}
\end{center} 
\caption{The photon distribution of a squeezed-thermal 
state as reconstructed by the ML estimation of the corresponding 
Wigner function (left histogram for the theoretical values, right
histogram for the reconstructed values). 
Number of data samples $N=50000$, quantum
efficiency $\eta =80\%$, number of thermal photons $n_{th}=0.1$,
number of squeezing photons $n_{sq}=3$. The statistical error affects
the third decimal digit, and  it is not visible in the scale of the
plot.} \label{f:sqth}
\end{figure}

 
In order to check the reliability of the state reconstruction we 
performed a set of Monte Carlo simulated experiments, starting from
homodyne measurements with quantum efficiencies in the range 
$\eta=70-90 \%$. For states with average photons in the range
$n_{th}<3$, $n_{coh}<5$, and $n_{sq}<3$ and for data samples of size
of the order $N=10^4-10 ^5$ we always found a reconstructed state very
close to the theoretical one. \footnote{As a global measure of the goodness 
of the reconstruction one should consider the normalized overlap 
$\cal O$ between the theoretical and the estimated state
$$ {\cal O}=\frac{\hbox{Tr}[\hat \varrho \,\hat \varrho _{\sc ml}]}{\sqrt
{\hbox{Tr}[\hat \varrho ^2]\,\hbox{Tr}[\hat \varrho _{\sc ml} ^2]}},$$ 
which is unit ${\cal O}=1$ if and only if  $\hat \varrho =\hat 
\varrho _{\sc ml}$.} 
As a matter of fact, the quality of the state reconstruction is good
enough that other physical quantities that are theoretically evaluated from
the experimental values of $\Delta _{\sc ml}, \kappa_{\sc ml}, a_{\sc ml}$ and 
$b_{\sc ml}$ are inferred very precisely. 
For example, we evaluated the photon number probability of a 
squeezed thermal state, which is given by the integral
\begin{eqnarray}
\langle n|\hat \varrho |n\rangle =\int _{0}^{2\pi}
\frac {d\phi}{2\pi} \frac{[C(\phi,n_{th},\kappa)-1]^n}
{C(\phi,n_{th},\kappa)^{n+1}}\;,
\end{eqnarray}
with $C(\phi,n_{th},\kappa)=(n_{th}+\frac
12)(\kappa ^{-1}\sin^2\phi+\kappa \cos^2\phi)+\frac 12$.  The comparison
of the theoretical and the experimental results for a state with
$n_{th}=0.1$ and $n_{sq}=3$ is reported in Fig. \ref{f:sqth}. The
statistical error of the reconstructed number probability affects the
third decimal digit, and is not visible on the scale of the plot.
\par As a further development, we mention that the ML estimation of Gaussian 
Wigner functions also provides a technique to estimate the coupling constants
of quadratic Hamiltonians of the form
\begin{eqnarray}
\hat H=\alpha a+\alpha ^* a^\dag + \varphi a^\dag a +\frac 12 \xi a^2+\frac
12 \xi^*a^{\dag 2}\;.\label{ham}
\end{eqnarray}
Hamiltonians like that in Eq. (\ref{ham}) 
describes the interaction of light modes
in active optical medium characterized by a second order susceptibility
tensor. Actually, the unitary evolution operator 
$\hat U=e^{-i\hat H t}$ preserves the
Gaussian form of an input state with Gaussian Wigner function, and therefore
one can use a Gaussian state to probe and characterize an
optical device.
\par\subsection{Absolute estimation of the quantum efficiency}
The operation of a photodetector is, in principle, very simple: 
each photon ionizes an atom, and the resulting charge is amplified 
to produce a measurable pulse. In practice, however, available 
photodetectors are usually characterized by a quantum efficiency 
lower than unity, which means that only a fraction of the incoming 
photons lead to an electric pulse, and ultimately to a "count". 
We may distinguish two main classes of photodetectors. In the first 
we have detectors where the resulting current
proportional to the incoming photon flux: in this case we have a 
linear photodetector. For example, this is the case of the high flux
photodetectors used in homodyne detection. In the second class we have
photodetectors operating at very low intensities, which resort to 
avalanche process in order to transform a single ionization event 
into a recordable pulse. This implies that one cannot discriminate between 
a single photon or many photons as the outcomes from such detectors 
are either a "click", corresponding to any number of photons, or 
"nothing" which means that no photons have been revealed. \par
Conventional 
characterization of photodetectors resorts to prepare a reference state with 
known intensity, and then measuring which fraction of the signal is 
actually revealed. This unavoidably leads to rather poor performances 
when applied in the relevant regime of quantum signals. Detection losses, 
in facts, distorce the whole probability distribution of the quantity 
being measured, not only the average value. Moreover, we need the accurate 
knowledge of the quantum state of the reference signal. In the following, 
we apply the ML principle to the absolute estimation of the quantum 
efficiency of both linear and avalanche photodetectors. We show that, 
along with the reliable characterization of quantum signals, 
ML method is an effective and statistically efficient tool 
for characterizing the response of a photodetector to low-intensity 
and/or nonclassical states. 
\par
\begin{figure}[h]\vspace{-5pt}\begin{tabular}{ccc} 
\psfig{file=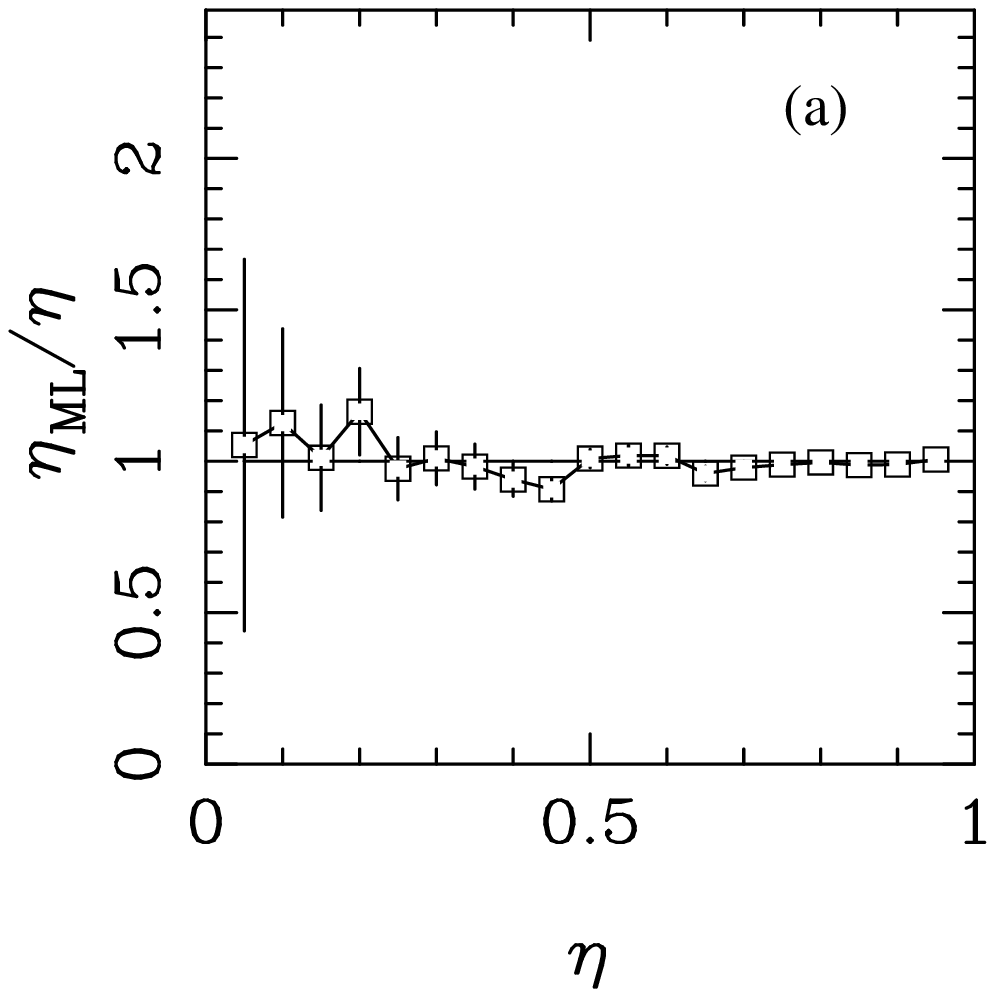,width=45mm} & & 
\psfig{file=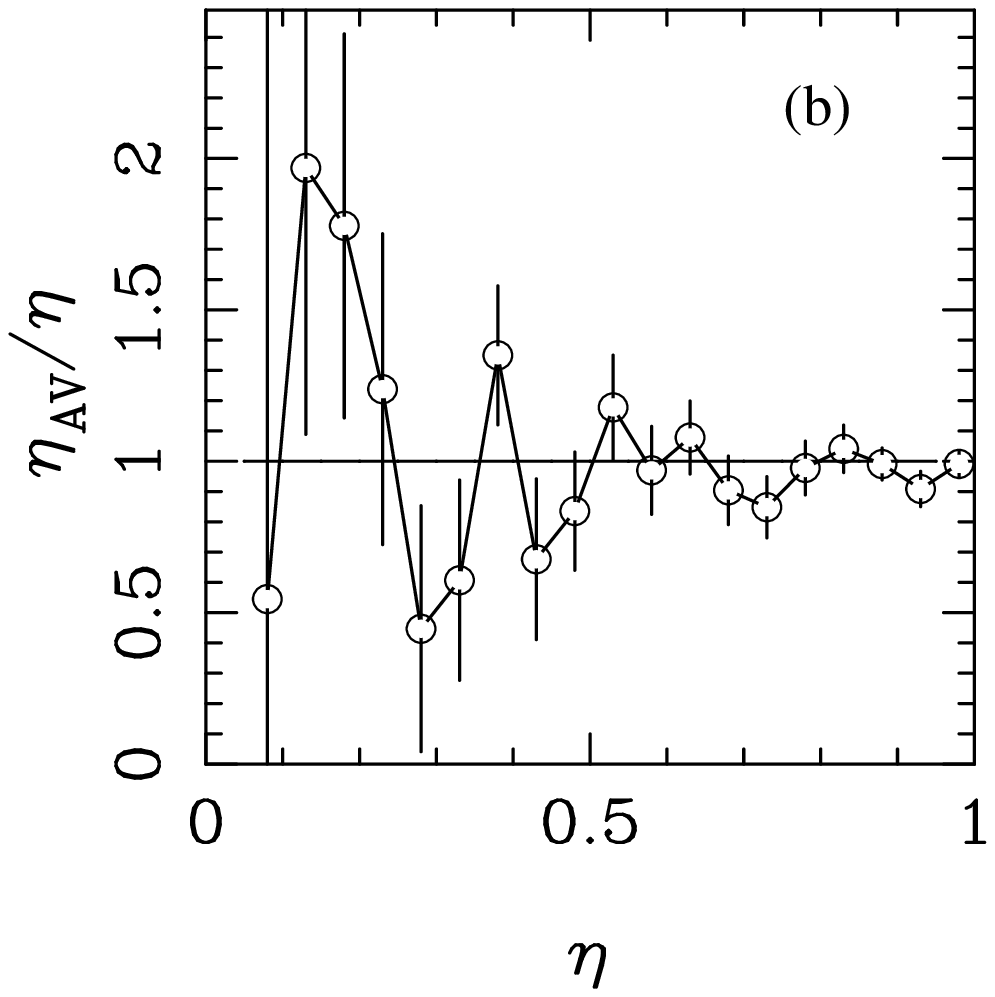,width=45mm} \end{tabular}
\caption{Quantum efficiency of linear photodetectors: ML estimation through 
homodyne detection on a squeezed state. The plots report the
ratio between the estimated value of the quantum efficiency and the
true value, as a function of the true value. 
On the left the results from maximum-likelihood method; on the right by 
the "naive" average-value method. Homodyne sample: $2500$ data. Reference state: 
a squeezed state with mean number of photons $n=1$ and squeezing fraction 
$\gamma = 99 \%$ (nearly a squeezed vacuum).}
\label{f:eta} \end{figure}

 Let us first study the case of linear
photodetectors. As a reference state we consider a squeezed-coherent
state, measured by homodyne detection.  The effect of  non-unit
quantum efficiency $\eta$ on the probability distribution of homodyne
detection is twofold. We have both a rescaling of the mean value and a
broadening of the distribution.  For a squeezed state $|x_0,r\rangle =
D (x_0) S(r) |0\rangle$ with the direction of squeezing
parallel to the signal phase and to the phase of the homodyne
detection (without loss of generality we set this phase equal to zero
and $x_0,r>0$ ) we have \cite{paul}
\begin{eqnarray}
p_\eta (x) = \frac1{\sqrt{2\pi\Delta^2}} 
\exp\left[-\frac{(x-\eta x_0)^2}{2\Delta^2}\right]\qquad
\Delta^2 = \frac14  \left(e^{-2r}+1-\eta\right)\label{pheta}\;.
\end{eqnarray}
The total number of photons of the state is given by $ n=x_0^2+\sinh^2 r $, 
whereas the squeezing fraction is
defined as $\gamma = \sinh^2r /n$.  Apart from an irrelevant constant, the 
log-likelihood function can be written as
\begin{eqnarray}
- L(\eta )= \log \Delta^2 + \frac{1}{\Delta^2}\left(\overline{x^2}+\eta x_0^2 -
2 \eta x_0 \overline{x}\right)
\label{loglikH}\;.
\end{eqnarray}
The resulting MLE is thus given by
\begin{eqnarray}
\eta_{\sc ml} = 1+ e^{-2r} + \frac1{x_0^2} \left\{
1-\sqrt{1+64x_0^2\left[\overline{x^2}+(1+e^{-2r})(x_0-2\overline{x}
+x_0e^{-2r})x_0\right]}\right\}
\nonumber
\end{eqnarray}
A set of Monte Carlo simulated
experiments confirmed that the Cram\'er-Rao bound is attained.  The
performances of the ML estimation can be compared to the "naive"
estimation based only on the measurement of the mean value, i.e.
$\eta_{\sc av} = \overline{x}/x_0$. We expect the naive method to be less
efficient, since the quantum efficiency not only rescales the mean
value, but also spreads the variance of the homodyne distribution in
Eq. (\ref{pheta}).  In Fig. \ref{f:eta}, on the basis of a Monte Carlo
simulated experiment, we compare the ML and the average-value methods
in estimating the quantum efficiency through homodyne detection on a
squeezed state.  The advantages of ML method are apparent, especially
for the estimation of low values of $\eta$. 
\par Let us now consider 
avalanche photodetectors, which perform the {\sf ON/OFF} measurement 
described by the two-value POM
\begin{eqnarray}
\hat \Pi_{\sc off} =  \sum_{p=0}^{\infty} (1-\eta)^p \: |p \rangle\langle p|
\qquad  \hat \Pi_{\sc on} = {\bf I} - \hat \Pi_{\sc off}
\label{yesno}\;, 
\end{eqnarray}
where ${\bf I}$ denotes the identity operator.  With avalanche
photodetectors we have only two possible outcomes: "click" or "no
clicks" which we denote by "1" and "0" respectively. The
log-likelihood function is given by
\begin{eqnarray}
L(\eta) = (N - N_c) \log P_0 (\eta) + N_c \log [1-P_0(\eta)]
\label{loglikyn}\;,
\end{eqnarray}
where $P_0 (\eta)=\hbox{Tr} [\hat \varrho  \hat \Pi_{\sc off}]$ is the
probability of having no clicks for the reference state described by
the density matrix $\varrho$, $N$ is the total number of
measurements, and $N_c$ is the number of events leading to a click.  The
maximum of $L(\eta)$, i.e. the MLE for the quantum efficiency,
satisfies the equation
\begin{eqnarray}
P_0 (\eta_{\sc ml}) = 1 - \frac{N_c}{N}
\label{MLyn}\;,
\end{eqnarray}
whose solution, of course, depends on the choice of the reference state.  The
optimal choice would be using single-photon states as a
reference. In this case, we have the trivial result $\eta_{\sc ml}=
N_c/N$. However, single-photon state are not easy to prepare 
and generally one would like to test $\eta $ for coherent pulses
$|\alpha\rangle$. In this case, we have $P_0(\eta)= \exp (-|\alpha
|^2\eta)$ and
\begin{eqnarray}
\eta_{\sc ml} = - \frac1{|\alpha |^2} \log \left( 1 - \frac{N_c}{N}\right)
\label{MLEyn}\;.
\end{eqnarray}
The Fisher information is given by
\begin{eqnarray}
F = \left(\frac{\partial P_0 }{\partial\eta}\right)^2 \frac1{P_0} + \left(
\frac{\partial P_1}{\partial\eta}\right)^2 \frac1{P_1}  
= \frac1{P_0(1-P_0)}\left(\frac{\partial P_0 }{\partial\eta}\right)^2 
\label{Fisheryn}\;,
\end{eqnarray}
and therefore, for a weak coherent reference one has  
\begin{eqnarray}
F = \frac{\eta^2}{e^{\eta |\alpha|^2}-1}\simeq \frac\eta{|\alpha|^2} 
\qquad\sigma_\eta &\simeq& \frac{|\alpha|}{\sqrt{\eta N}}
\label{confyn}\:.
\end{eqnarray}
\par\section{Summary and conclusions}
In this paper we reviewed the application of the maximum likelihood 
principle to the reconstruction of the density matrix of a generic 
quantum system \cite{maxlik}, as well as to the estimation 
of relevant parameters 
in quantum optics \cite{parlik}. In all cases, the resulting 
reconstruction algorithm is statistically efficient, and provides 
the reliable estimation of the quantity of interest using much smaller 
data samples compared to conventional methods. In particular, the ML 
estimation of the density matrix allows one to incorporate the natural 
physical constraints we have on the quantum state, thus leading to a 
substantial reduction of statistical fluctuations. For quantum-optical 
parameters, the ML approach provides efficient estimation 
schemes based on feasible measurements like homodyne detection. 
The resulting procedures lead to substantial improvement over conventional
methods and are of technological interest. 
\par\section*{Acknowledgement}
This work has been supported by INFM through the project PRA-97-CAT.
The ML estimation algorithm for quantum state ( i.e. the subject of 
section 3) has been developed by the authors in collaboration with 
Konrad Banaszek during his stay at the INFM unit of Pavia. MGAP 
thanks K.B. and Zdenek Hradil for interesting discussions.
\par

\begin{thebibliography}{99}
\bibitem{not} In terms of $\hat{\varrho}$-eigenvectors
$| \psi_\mu \rangle $, one has  as $ \hat{\varrho} = \sum_{\mu } y_{\mu}^{2} |
\psi_\mu \rangle \langle \psi_\mu | $. The 
maximum likelihood condition $\partial L/\partial y_{\nu} = 0$ gives 
$\lambda y_{\nu} = \sum_{i=1}^{N} [y_\nu \langle \psi_\nu | \hat{\cal
F}_i | \psi_\nu \rangle/ \hbox{Tr}(\hat\varrho \hat{\cal F}_i)]$. 
After multiplication by $y_{\nu}$ and summation over $\nu $, one
obtains $\lambda = N$.
\end{thebibliography}
\end{document}